# Aspects of the Scattering and Impedance Properties of Chaotic Microwave Cavities

Sameer Hemmady[1,2,3], Xing Zheng[3], Thomas M. Antonsen Jr.[1,3], Edward Ott[1,3] and Steven M. Anlage[1,2]
*Department of Physics, University of Maryland, College Park, MD 20742-4111 USA*

**Abstract:**
We consider the statistics of the impedance $Z$ of a chaotic microwave cavity coupled to a single port. We remove the non-universal effects of the coupling from the experimental $Z$ data using the radiation impedance obtained directly from the experiments. We thus obtain the normalized impedance whose Probability Density Function (PDF) is predicted to be universal in that it depends only on the loss (quality factor) of the cavity. We find that impedance fluctuations decrease with increasing loss. The results apply to scattering measurements on any wave chaotic system.

**PACS# : 05.45.Mt, 03.65.Nk,11.55.-M,03.50.De,84.40.-X,84.40.Az**

## 1. Introduction:

Quantum chaotic scattering is a subject that first arose in the context of nuclear scattering [1]. It is now finding increasing applications in condensed matter and atomic physics to understand the properties of larger scale complicated quantum systems [2, 3, 4]. Quantum scattering is also of interest for understanding the universal statistical properties of complicated electromagnetic enclosures [5]. Here we are concerned with systems that display ray chaos in the limit of high quantum number, or wave energy. The quantities of interest in the corresponding wave/quantum system are the impedance ($Z$) and scattering ($S$) matrices. The system has a finite number of ports or scattering channels, and the $S$-matrix relates outgoing waves in terms of a linear combination of incoming waves on the system. The $Z$ matrix relates the total voltage at one port to a linear combination of the currents entering all of the ports. They are related by a simple bi-linear transformation $S = (Z + Z_o)^{-1}(Z - Z_o)$, where $Z_0$ is the characteristic impedance of the ports.

One successful approach to quantifying the statistical properties of quantum scattering systems is the Poisson Kernel (PK) [6-8]. The PK is a global approach to understanding the statistical properties of scattering systems. It describes the statistical properties of the scattering matrix $S$ in the presence of imperfect coupling, in terms of the mean value of S. The mean value $\bar{S}$ characterizes the imperfect coupling between the system and the exterior scattering channels, and can be approximately evaluated from data by taking the mean of a large amount of data on an ensemble of similar systems, or by energy averaging, or both. This approach has proven to be quite useful for describing microwave scattering data, for instance [9,10].

We have introduced a complementary new approach to quantum scattering that makes use of different conceptual pieces to achieve a similar outcome. Our approach is to directly characterize the non-ideal coupling between the outside world and the system through a deterministic quantity known in electromagnetism as the radiation impedance, $Z_{Rad}$. One can then define normalized impedance (z) and scattering (s) matrices that directly reveal the universal fluctuating properties of the scattering system [11-13]. More explicitly, the radiation impedance $Z_{Rad} = R_{Rad} + iX_{Rad}$ of each channel is determined in a separate measurement and combined with the cavity impedance $Z = R + iX$ to create a normalized impedance matrix $z$ as;
$z = \frac{R}{R_{Rad}} + i\frac{X - X_{Rad}}{R_{Rad}}$. This normalized impedance matrix has statistical properties that are intrinsic to the scattering system and independent of the coupling. A number of experiments have been performed to test the universality of the normalized $z$ [14, 15]. The normalized impedance approach has also been employed by the Warszawa group to examine their data on quantum graph microwave analogs [16].

---
[1] Also with the Department of Electrical and Computer Engineering.
[2] Also with the Center for Superconductivity Research.
[3] Also with the Institute for Research in Electronics and Applied Physics.

## 2. Experiment:

Our experimental setup consists of an air-filled quarter bow-tie chaotic cavity (Fig.1(a)) which acts as a two dimensional resonator below about 19 GHz [17]. Ray trajectories in a closed billiard of this shape are known to be chaotic. This cavity has previously been used for the study of the eigenvalue spacing statistics [18] and eigenfunction statistics [19, 20] for a wave chaotic system. In order to investigate a scattering problem, we excite the cavity by means of a single coaxial probe whose exposed inner conductor, with a diameter (2a) extends from the top plate and makes electrical contact with the bottom plate of the cavity (Fig.1(b)). In this paper we study the properties of the cavity over a frequency range of 6 - 12 GHz, where the spacing between two adjacent resonances is on the order of 25 – 30 MHz.

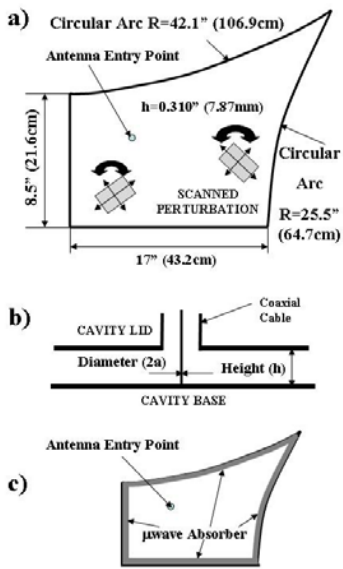

Fig.1: (a) The physical dimensions of the quarter bow-tie chaotic microwave resonator are shown along with the position of the single coupling port. Two metallic perturbations are systematically scanned and rotated throughout the entire volume of the cavity to generate the cavity ensemble. (b) The details of the coupling port (antenna) and cavity height $h$ are shown in cross section. (c) The implementation of the radiation case is shown, in which commercial microwave absorber is used to line the inner walls of the cavity to minimize reflections.

As in the numerical experiments in Refs. [12, 13], our experiment involves a two-step procedure. The first step is to collect an ensemble of cavity scattering coefficients $S$ over the frequency range of interest. Ensemble averaging is realized by using two rectangular metallic perturbations which are systematically scanned and rotated throughout the volume of the cavity (Fig.1(a)). Each configuration of the perturbers within the cavity volume results in a different measured value of $S$. The perturbers are kept far enough from the antenna so as not to alter its near-field characteristics. In total, one hundred different configurations are measured, resulting in an ensemble of 800,000 $S$ values. We refer to this step as the "cavity case".

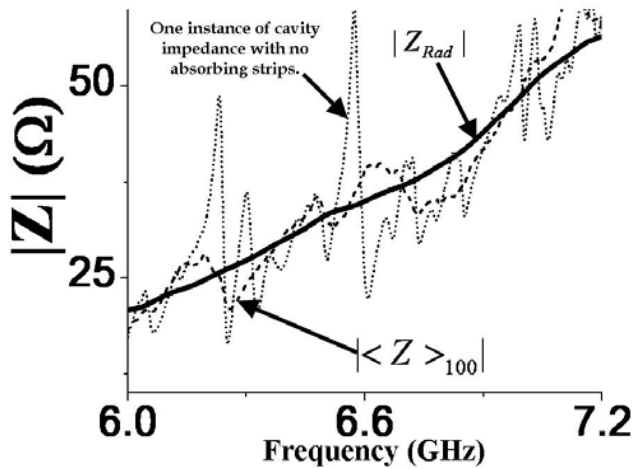

Fig. 2: The magnitude of the cavity impedance with no absorbing strips is shown as a function of frequency. The dots indicate a single rendition of the cavity impedance and perturbations. The dashed line is the magnitude of the complex cavity impedance obtained after ensemble averaging over 100 different perturbation positions within the cavity. The solid line is the magnitude of the measured radiation impedance for the same antenna and coupling detail as shown in Fig.1 (b). Note that even after 100 renditions of the perturbers within the cavity, $|<Z>_{100}|$ is still a poor approximation to $|Z_{Rad}|$.

The second step, referred to as the "radiation case", involves obtaining the scattering coefficient for the excitation port when waves enter the cavity but do not return to the port. In the experiment, this condition is realized by removing the perturbers and lining the side walls of the cavity with commercial microwave absorber (ARC Tech DD10017D) which provides about 25 dB of reflection loss between 6 and 12 GHz (Fig.1.(c)). We measure the radiation scattering coefficient $S_{Rad}$ for the cavity, approximating the situation where the side walls are moved to infinity. In this case $S_{Rad}$ does not depend on the chaotic ray trajectories of the cavity, and thus gives a deterministic (i.e. non-statistical) characterization of the coupling independent of the chaotic system.

Having measured the cavity $S$ and $S_{Rad}$, we then transform these quantities into the corresponding cavity and radiation impedances ($Z$ and $Z_{Rad}$) and determine the normalized impedance $z$ as discussed above.

In order to test the validity of the theory for systems with varying loss, we create different "cavity cases" with different degrees of loss. Loss is controlled and parameterized by placing 15.2 cm-long strips of microwave absorber along the inner walls of the cavity. These strips cover the side walls from the bottom to top lids of the cavity. We thus generate different loss scenarios by increasing the number of 15.2 cm -long absorber strips placed along the inner cavity walls and define the absorber perimeter ratio $\alpha$ as the ratio of absorber length to the total cavity perimeter (147.3 cm).

### 3. Data:

We first examine the degree to which ensemble averaging to estimate $\bar{S}$ and $\bar{Z}$, as employed in the Poisson Kernel, can reproduce the radiation cases $S_{Rad}$ and $Z_{Rad}$. Fig. 2 shows typical data for the magnitude of the cavity impedance versus frequency for several cases. The dots show the cavity impedance for one particular rendition of the cavity and its perturbers in the case of no added absorber. The dashed line shows the result of averaging the complex impedance of 100 renditions of the cavity. The thick solid line is the measured radiation impedance $|Z_{Rad}|$, which should be equivalent to the mean of the cavity impedance $\bar{Z}$. It is clear that even after averaging the properties of 100 cavities in the ensemble, the mean value of measured $Z$ has not yet approached the radiation case. This demonstrates the importance of obtaining very high quality statistics before the Poisson Kernel can be used on real data. It also illustrates the relative ease with which the radiation impedance can be used to characterize the non-ideal coupling of real wave chaotic systems.

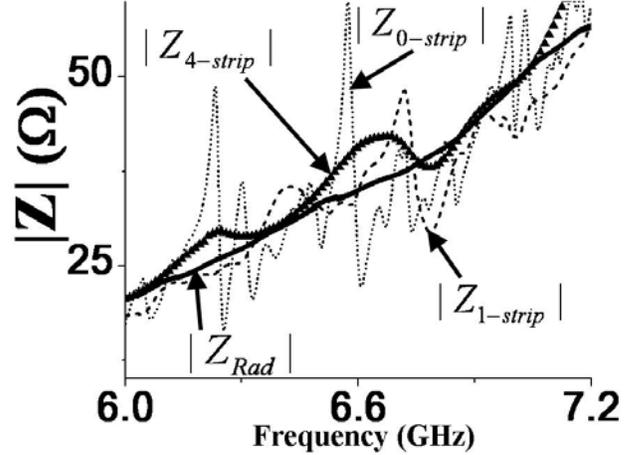

Fig. 3: The magnitude of a single rendition of the cavity impedance with (0-dots, 1-dashed line, 4-solid triangles) absorbing strips is shown as a function of frequency. The solid line is the magnitude of the measured radiation impedance for the same antenna and coupling detail as shown in Fig. 1(b). As losses within the cavity increase, the cavity resonances are dampened out and the measured cavity impedance approaches the radiation impedance.

We next examine the dependence of impedance statistics on the global control parameter $k^2/(\Delta k_n^2 Q)$. This parameter depends on the frequency, the volume of the scattering system (through the resulting mode density $\Delta k_n^2$), and on the losses (parameterized by the scatterer quality factor $Q$). It determines the probability density functions (pdf) for the real and imaginary parts of the normalized impedance $z$, as well as the pdf of $|s|$. One can determine the value of $k^2/(\Delta k_n^2 Q)$ from the variance of the $\text{Re}[z]$ and $\text{Im}[z]$ pdfs, as shown in [14], since the variance $\sigma^2 = (1/\pi)/(k^2/(\Delta k_n^2 Q))$ for systems with time-reversal symmetry.

Fig. 3 demonstrates how the cavity impedance evolves with increasing loss. Shown are impedance magnitude data versus frequency for 3 cavities with different numbers of microwave absorber strips inside (0, 1, or 4), but otherwise identical. These data sets are for a single rendition of the cavity. Also shown is the measured radiation impedance magnitude for the same antenna. As losses increase, the fluctuations in |Z| clearly decrease, and approach the radiation case. This qualitative observation is substantiated by quantitative tests of the Re[z] and Im[z] pdf variances, and their dependence on system loss [14].

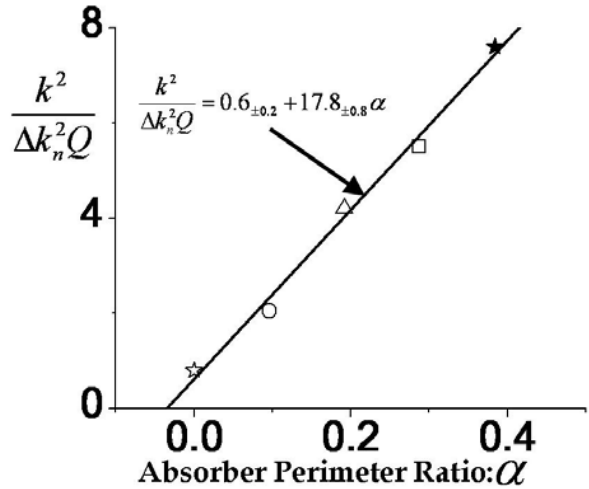

**Fig. 4:** The relationship between the loss parameter $k^2/(\Delta k_n^2 Q)$ and the absorber perimeter ratio ($\alpha$) is shown between 7.2 and 8.4 GHz. The symbols represent (hollow star-0 ; hollow circle- 1; hollow triangle- 2; hollow square- 3; solid star- 4 ) absorbing strips within the cavity. The best linear fit to all the data points is shown as the solid line. The x-intercept of this line indicates the $\alpha$ required to make a loss-less cavity have the same $k^2/(\Delta k_n^2 Q)$ as the empty experimental cavity of Fig. 1(a).

Figure 4 further examines the dependence of the experimentally determined value of $k^2/(\Delta k_n^2 Q)$ versus the number of absorber strips placed along the periphery of the cavity walls. The $k^2/(\Delta k_n^2 Q)$ values were determined by the variances of the Re[$z$] and Im[$z$] pdfs. Fig. 4 shows a clear linear relationship of $k^2/(\Delta k_n^2 Q)$ on the absorber perimeter ratio. This is expected because $1/Q$ is proportional to the dissipated power in the cavity, which scales with the amount of microwave absorber placed in the cavity. A linear fit of the data is quite accurate and shows a zero-crossing for $k^2/(\Delta k_n^2 Q)$ at $\alpha = -0.035$. This suggests that the empty cavity losses correspond to covering the walls of a perfectly conducting cavity with 3.5% coverage of microwave absorber.

### 4. Discussion and Conclusions:

Our work has illustrated the benefits of using the impedance, rather than scattering matrix, to determine the universal scattering properties of wave chaotic systems connected to the outside world. The measurement of radiation impedance to characterize the non-ideal coupling to the system is a very simple and powerful tool, and is more reliable than an average statistical quantity. The evolution of wave chaotic scattering systems with internal losses has been illustrated in this paper. The impedance approach also reveals other universal properties of the cavity, such as the generalized variance ratio, related to the Hauser-Feshbach relation [21].

### Acknowledgements:

We acknowledge useful discussions with R. Prange and S. Fishman, as well as comments from Y. Fyodorov, D.V. Savin and P. Brouwer. This work was supported by the DOD MURI for the study of microwave effects under AFOSR Grant F496200110374 and an AFOSR DURIP Grant FA95500410295.